\documentclass{article}

\usepackage{PRIMEarxiv}

\usepackage[utf8]{inputenc} 
\usepackage[T1]{fontenc}    
\usepackage{hyperref}       
\usepackage{url}            
\usepackage{booktabs}       
\usepackage{amsfonts}       
\usepackage{nicefrac}       
\usepackage{microtype}      
\usepackage{lipsum}
\usepackage{fancyhdr}       
\usepackage{graphicx}       
\graphicspath{{media/}}     

\usepackage{xcolor}
\usepackage{amsmath}

\title{End-to-end differentiable design of geometric waveguide displays}

\author{  
  {\rm Xinge Yang$^{1}$ \quad Zhaocheng Liu$^{2, *}$ \quad Zhaoyu Nie$^{2}$ \quad Qingyuan Fan$^{2}$} \\ {Zhimin Shi$^{2}$ \quad Jim Bonar$^{2}$ \quad Wolfgang Heidrich$^{1, *}$} \\[8pt]  
  KAUST$^{1}$, Meta$^{2}$, Corresponding author$^{*}$ \\
}  

\begin{document}
\maketitle


\begin{abstract}
Geometric waveguides are a promising architecture for optical see-through augmented reality displays, but their performance is severely bottlenecked by the difficulty of jointly optimizing non-sequential light transport and polarization-dependent multilayer thin-film coatings. Here we present the first end-to-end differentiable optimization framework for geometric waveguide that couples non-sequential Monte Carlo polarization ray tracing with a differentiable transfer-matrix thin-film solver. A differentiable Monte Carlo ray tracer avoids the exponential growth of deterministic ray splitting while enabling gradients backpropagation from eyebox metrics to design parameters. With memory-saving strategies, we optimize more than one thousand layer-thickness parameters and billions of non-sequential ray-surface intersections on a single multi-GPU workstation. Automated layer pruning is achieved by starting from over-parameterized stacks and driving redundant layers to zero thickness under discrete manufacturability constraints, effectively performing topology optimization to discover optimal coating structures. On a representative design, starting from random initialization within thickness bounds, our method increases light efficiency from 4.1\% to 33.5\% and improves eyebox and FoV uniformity by $\sim$17$\times$ and $\sim$11$\times$, respectively. Furthermore, we jointly optimize the waveguide and an image preprocessing network to improve perceived image quality. Our framework not only enables system-level, high-dimensional coating optimization inside the waveguide, but also expands the scope of differentiable optics for next-generation optical design.
\end{abstract}

\section*{Introduction}

Augmented reality (AR) overlays digital content onto the real world and motivates compact, lightweight optical combiners for near-eye displays~\cite{lee2020foveated,Xiong_2021,kress2013review,shi2025flat}. Existing optical see-through architectures span birdbath-type combiners~\cite{hua20143d,rolland2000wide}, which can offer high image quality but remain bulky and reduce transparency, and retinal projection systems~\cite{Jang_2017,Lee_2018,Tseng_2024}, which can be power-efficient but require precise alignment and provide limited eyebox. Waveguide displays~\cite{rolland2024waveguide,Weng_2025} provide an attractive alternative by using a thin transparent slab to guide light via total internal reflection (TIR) and replicate the exit pupil without bulky optics, representing a promising architecture for next-generation AR displays.

Waveguide displays commonly include diffractive/holographic waveguides~\cite{Gopakumar_2024,Jang_2024, Moharam_1982,Miller_1997,Ni_2022,weng2023high,choi2025synthetic} and geometric waveguides (GWGs)~\cite{Danziger2021,Cheng_2022,gu2018design,Xu_2019}. Diffractive approaches leverage wavelength-dependent diffraction to couple light in and out of the slab and often require additional design effort to manage colour and angular sensitivity. GWGs instead use partially reflective mirror arrays (PRMAs) to redirect and extract light using geometric optics. Because coupling is achieved by reflection rather than diffraction, GWGs can preserve spectral content and offer a practical route to high image quality with large eyebox. Reflective coupling can also deliver high light efficiency, making GWGs attractive for bright near-eye displays.

Despite this promise, GWG design remains challenging because performance depends on non-sequential light transport through many partially reflective interactions and on polarization-dependent multilayer coatings. First, accurate simulation of non-sequential ray paths is often performed by ray splitting into reflected and transmitted branches at each PRMA interaction, which can lead to exponential growth in ray count and high computational cost. Second, existing workflows typically decouple geometry and coating design~\cite{Cheng_2022,Ruan_2023,Xu_2019}, making iterative refinement slow and hindering system-level optimization. Critically, these decoupled approaches often fail to account for polarization effects during the design phase, leading to significant performance degradation in fabricated prototypes. Third, optical optimization is commonly tackled with sampling-based strategies (for example genetic algorithms) or finite-difference gradient estimation. The former is typically inefficient and often trapped in local minima within high-dimensional parameter spaces, while the latter is computationally expensive for multilayer stacks. Although differentiable optics has enabled gradient-based design for a range of optical systems~\cite{Yang_2024,Wang_2022,Tseng_2024,yang2024hybrid}, its application to GWGs, with coupled non-sequential transport and thin-film polarization physics, remains unexplored to our knowledge. Together, these limitations make GWG design time-consuming and computationally expensive, often requiring cluster-scale resources and multi-day optimization to meet targets in light efficiency and uniformity.

Here we enable scalable gradient-based optimization of GWG coatings with an end-to-end differentiable simulation spanning PRMA geometry and multilayer thin films. Specifically, we (i) introduce differentiable non-sequential Monte Carlo polarization ray tracing, in which reflection/transmission events are sampled probabilistically at each partially reflective mirror. We (ii) integrate a differentiable thin-film solver based on the transfer matrix method~\cite{heavens1960optical} to capture coating-induced polarization effects. This end-to-end differentiable formulation enables efficient simulation and backpropagates gradients directly from eyebox metrics to design parameters. Moreover, we (iii) combine memory-saving strategies to support optimization on a single multi-GPU workstation, and introduce a discrete optimization strategy that automatically prunes unnecessary coating layers by driving them to zero thickness. Taken together, we turn GWG coating design from decoupled, mirror-by-mirror tuning into a tractable, system-level gradient optimization problem.

\begin{figure}
    \centering
    \includegraphics[width=\textwidth]{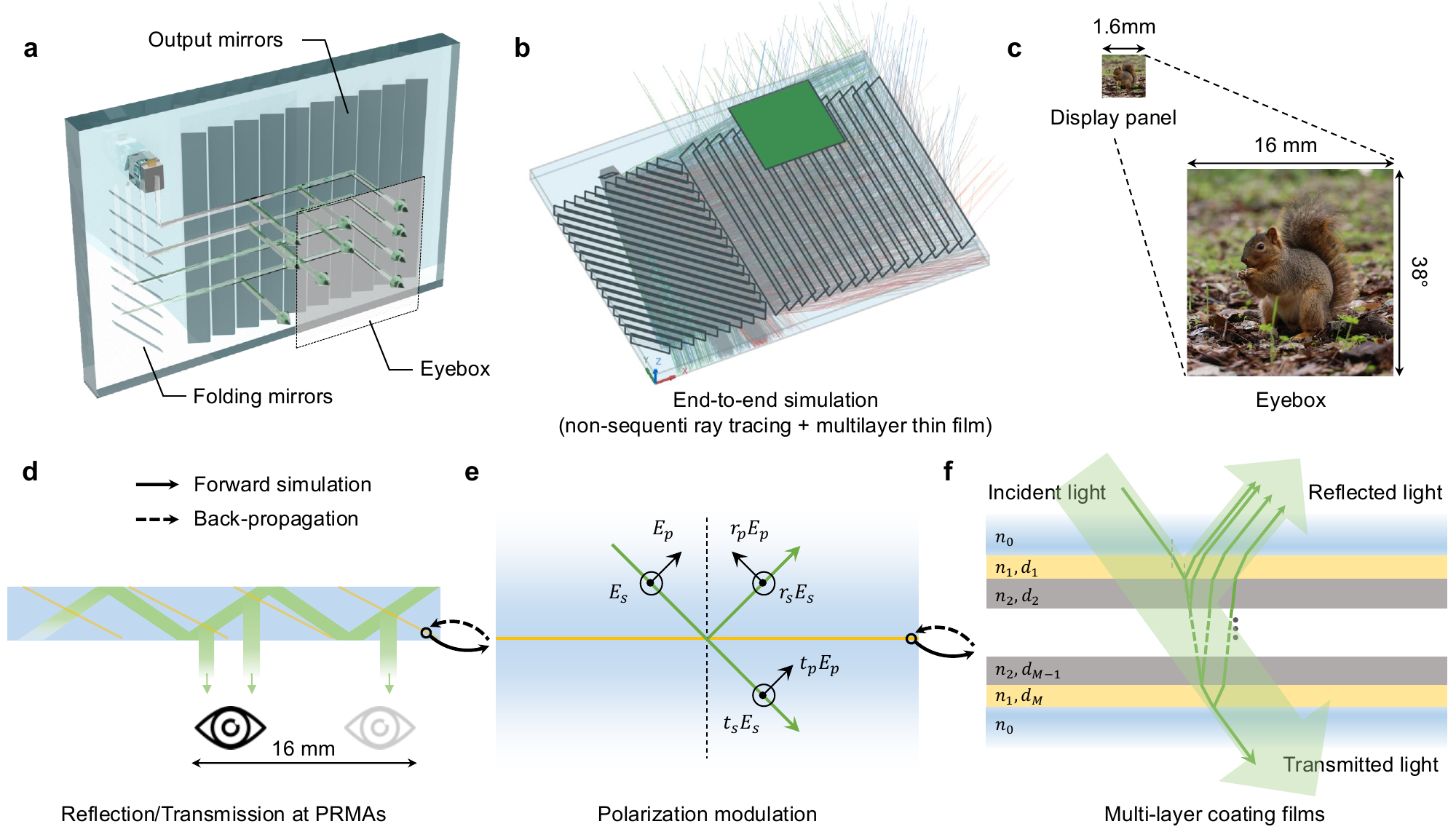}
    \caption{\textbf{Illustration of geometric waveguide architecture and our proposed differentiable optimization.} \textbf{a} The GWG employs partially reflective mirror arrays to redirect light from the display engine to the user's eye pupil. The exit pupil is replicated and expanded relative to the input pupil (display panel). \textbf{b} We establish an end-to-end differentiable simulation spanning PRMA geometry and multilayer coatings on each mirror, enabling system-level optimization. \textbf{c} A representative GWG achieves a 100$\times$ pupil expansion, increasing from a 1.6\,mm $\times$ 1.6\,mm display panel to 16\,mm $\times$ 16\,mm at the eyebox with a FoV of 38$^\circ$ $\times$ 38$^\circ$. \textbf{d} At each partially reflective mirror, light is either reflected or transmitted. We use differentiable non-sequential Monte Carlo ray tracing to simulate these paths within the GWG. At the eyebox region, output mirror arrays couple light out across a large region. \textbf{e} Each mirror is coated with multilayer thin films that modulate polarization. \textbf{f} A differentiable transfer-matrix solver~\cite{heavens1960optical} computes effective Fresnel coefficients for multilayer coatings. In the forward pass, rays carry gradient information through the sampled paths. In backpropagation, gradients of the eyebox image with respect to film thicknesses are computed automatically, enabling gradient-based optimization.}
    \label{fig:gwg}
\end{figure}

To demonstrate these capabilities, we optimize a representative GWG architecture end-to-end. Using our differentiable Monte Carlo estimator and transfer-matrix thin-film model, we jointly optimize all PRMA coating stacks starting from random initialization within manufacturable thickness bounds. Despite the large scale of more than one thousand layer-thickness parameters and billions of non-sequential ray-surface intersections, the full optimization runs on a single multi-GPU workstation and converges automatically in hours. On this design, both light efficiency and uniformity (FoV and eyebox) are improved substantially. We cross-validate the simulator against deterministic ray splitting and a reference thin-film solver (Supplementary Note), and show faster convergence and better optima than a genetic-algorithm baseline. Finally, we extend the framework to system-level optical-digital co-design by jointly optimizing the waveguide and a neural image preprocessing network to compensate residual nonuniformity at the eyebox. 
\begin{figure}[t]
    \centering
    \includegraphics[width=0.95\textwidth]{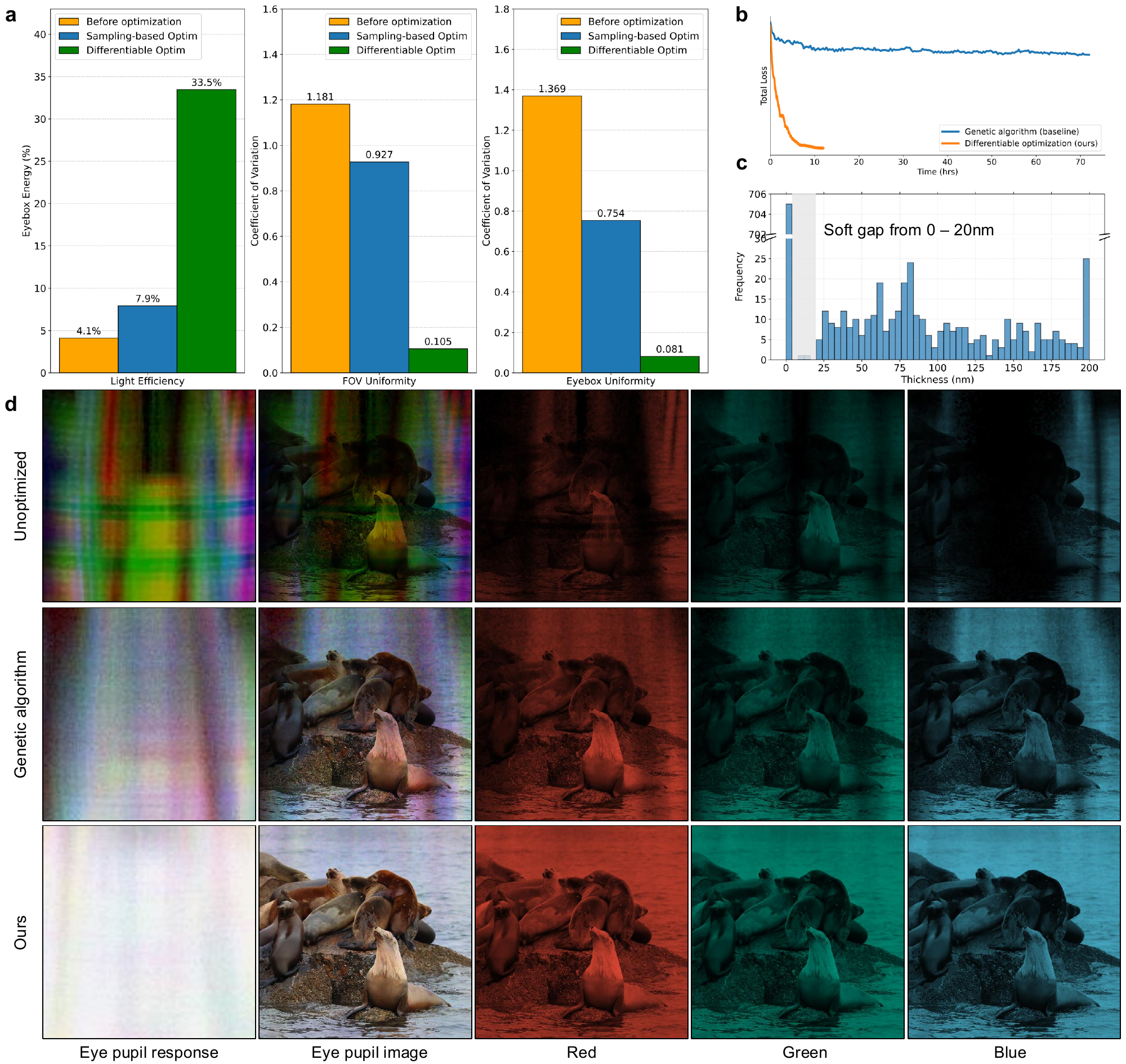}
    \caption{\textbf{Evaluation of end-to-end differentiable optimization for GWG coatings.} \textbf{a} Light efficiency and uniformity for the initial design, a genetic-algorithm baseline and our differentiable optimization, measured by eyebox throughput $\bar{I}$ (fraction of input power reaching the eyebox), $\text{CV}_{\text{FoV}}$ and $\text{CV}_{\text{eyebox}}$ (all metrics averaged over five Monte Carlo runs with different random seeds). \textbf{b} Loss curves for differentiable optimization and the genetic-algorithm baseline. Gradient-based optimization converges faster and reaches a lower loss. \textbf{c} Thickness distribution of the optimized multilayer stack. Several layers are driven towards zero thickness, indicating that they can be removed in the final design. The discrete optimization strategy supports an over-parameterized starting stack followed by pruning during optimization. \textbf{d} Simulated pupil response ($256 \times 256$) at the centre of the eyebox, displayed image and RGB channels ($512 \times 512$) for the initial design, the genetic algorithm and differentiable optimization. Differentiable optimization increases brightness and reduces FoV and eyebox non-uniformities.}
    \label{fig:results}
\end{figure}

\section*{Results}

We optimize all PRMA coating stacks jointly by differentiating through non-sequential Monte Carlo polarization transport and a transfer-matrix thin-film model. Starting from random initialization within thickness bounds and pruning an over-parameterized stack during optimization, we increase light efficiency by $\sim$8$\times$ and reduce non-uniformity across the FoV by $\sim$11$\times$ and eyebox by $\sim$17$\times$, respectively, outperforming a genetic-algorithm baseline (Fig.~\ref{fig:results}). We first detail the waveguide architecture and objective, then analyze the optimized stacks and pruning behaviour, and quantify the memory-saving strategies that enable optimization at scale. Finally, we present optical-digital co-design as a system-level extension.

\subsection*{System Architecture}

Figure~\ref{fig:gwg}\textbf{a} shows the GWG configuration considered. The waveguide uses a 1.7-mm-thick glass substrate (refractive index 1.9). An input mirror couples light from a 1.6~mm $\times$ 1.6~mm display panel into the waveguide, where rays undergo multiple total internal reflections (TIR) before reaching the partially reflective mirror arrays (PRMAs). The GWG contains 30 folding mirrors for vertical pupil expansion and 16 output mirrors for horizontal expansion and out-coupling towards the eyebox (Fig.~\ref{fig:gwg}\textbf{b,c}). At each PRMA interaction, light is either reflected or transmitted (Fig.~\ref{fig:gwg}\textbf{d}), and rays propagate non-sequentially between mirrors and waveguide boundaries until they exit towards the eyebox. The eyebox is 16~mm $\times$ 16~mm at a 15~mm eye relief, corresponding to 10$\times$ horizontal and 10$\times$ vertical pupil expansion, and the FoV is $38^\circ \times 38^\circ$. Reflection and transmission are polarization-dependent and are controlled by multilayer coatings via effective Fresnel coefficients (Fig.~\ref{fig:gwg}\textbf{e,f}). Each mirror is coated with a 23-layer Ta$_2$O$_5$/SiO$_2$ stack, yielding more than one thousand optimizable layer thicknesses across all PRMAs. Thicknesses are constrained to 20-200~nm, with an additional mechanism that allows layers to be pruned by driving their thickness to 0.

We simulate light propagation with non-sequential Monte Carlo ray tracing and model coating-induced polarization effects with a thin-film solver (Fig.~\ref{fig:gwg}\textbf{b}). For each FoV sample, we emit parallel rays from the display panel and trace them through the waveguide. To model chromatic effects, we average three wavelengths per RGB channel (red: 620/660/700~nm; green: 500/530/560~nm; blue: 450/470/490~nm). Sampling across FoV angles yields a two-dimensional RGB image for each pupil position. We evaluate a 3$\times$3 grid of pupil positions within the eyebox and use the corresponding pupil images as the optimization objective (see Supplementary Note).

\subsection*{End-to-end Differentiable Coating Film Design}

The end-to-end differentiable model enables backpropagation from the eyebox image to the multilayer thickness parameters through the sampled ray paths. We optimize brightness and uniformity using the multi-objective loss
\begin{equation}
    \mathcal{L} = \mathcal{L}_{\text{bright}} + w_f \cdot \mathcal{L}_{\text{FoV}} + w_e \cdot \mathcal{L}_{\text{eyebox}},
    \label{eq:loss}
\end{equation}
where $\mathcal{L}_{\text{bright}} = -\bar{I}$ encourages brightness and $\bar{I}$ is the mean eyebox throughput (fraction of input power reaching the eyebox) averaged over pupil positions, FoV samples and colour channels. To quantify non-uniformity, we use the coefficient of variation (CV; standard deviation divided by mean), which is less scale-sensitive than variance and helps avoid convergence to near-zero intensity solutions. We set $\mathcal{L}_{\text{FoV}}=\text{CV}_{\text{FoV}}$ and $\mathcal{L}_{\text{eyebox}}=\text{CV}_{\text{eyebox}}$, and use weights $w_f=10.0$ and $w_e=3.0$. We run 1,000 optimization iterations with a $32 \times 32$ angular FoV grid and 10,000 rays per FoV; each ray is traced for up to 100 interactions.

Layer thicknesses are initialized from a Gaussian distribution centred at the midpoint of the thickness bounds. After differentiable optimization, eyebox throughput increases from 4.1\% to 33.5\% ($\sim$8$\times$). Uniformity also improves: $\text{CV}_{\text{FoV}}$ decreases from 1.181 to 0.105 ($\sim$11$\times$) and $\text{CV}_{\text{eyebox}}$ decreases from 1.369 to 0.081 ($\sim$17$\times$) (Fig.~\ref{fig:results}\textbf{a}). A sampling-based genetic-algorithm baseline is adopted for comparison and achieves 7.9\% eyebox throughput. To be conservative, we allowed the genetic algorithm six times the wall-clock compute (72 h versus 12 h) under identical forward-simulation settings. Gradient-based optimization converges faster and reaches a lower loss compared to the genetic-algorithm baseline, which struggles to explore the high-dimensional landscape effectively (Fig.~\ref{fig:results}\textbf{b}). Relative to this baseline, differentiable optimization achieves $\sim$4.2$\times$ higher $\bar{I}$ and reduces $\text{CV}_{\text{FoV}}$ and $\text{CV}_{\text{eyebox}}$ by $\sim$9$\times$ and $\sim$9$\times$, respectively. During optimization, some layers are driven towards zero thickness (Fig.~\ref{fig:results}\textbf{c}), enabling topological pruning from an over-parameterized starting stack. Under our discrete thickness strategy, we introduce a soft gap between 0 and 20~nm, reflecting a practical minimum thickness while still allowing layers to be eliminated. Figure~\ref{fig:results}\textbf{d} shows the simulated centre-pupil response ($512 \times 512$), displayed image and RGB channels for the initial design, the genetic algorithm and differentiable optimization. The initial design exhibits strong non-uniformities and dead regions in the displayed image, whereas differentiable optimization increases brightness and reduces FoV and eyebox non-uniformities. All reported metrics are averaged over five Monte Carlo runs with different random seeds. Additional evaluations are provided in the Supplementary Note.

\begin{figure}[t]
    \centering
    \includegraphics[width=\textwidth]{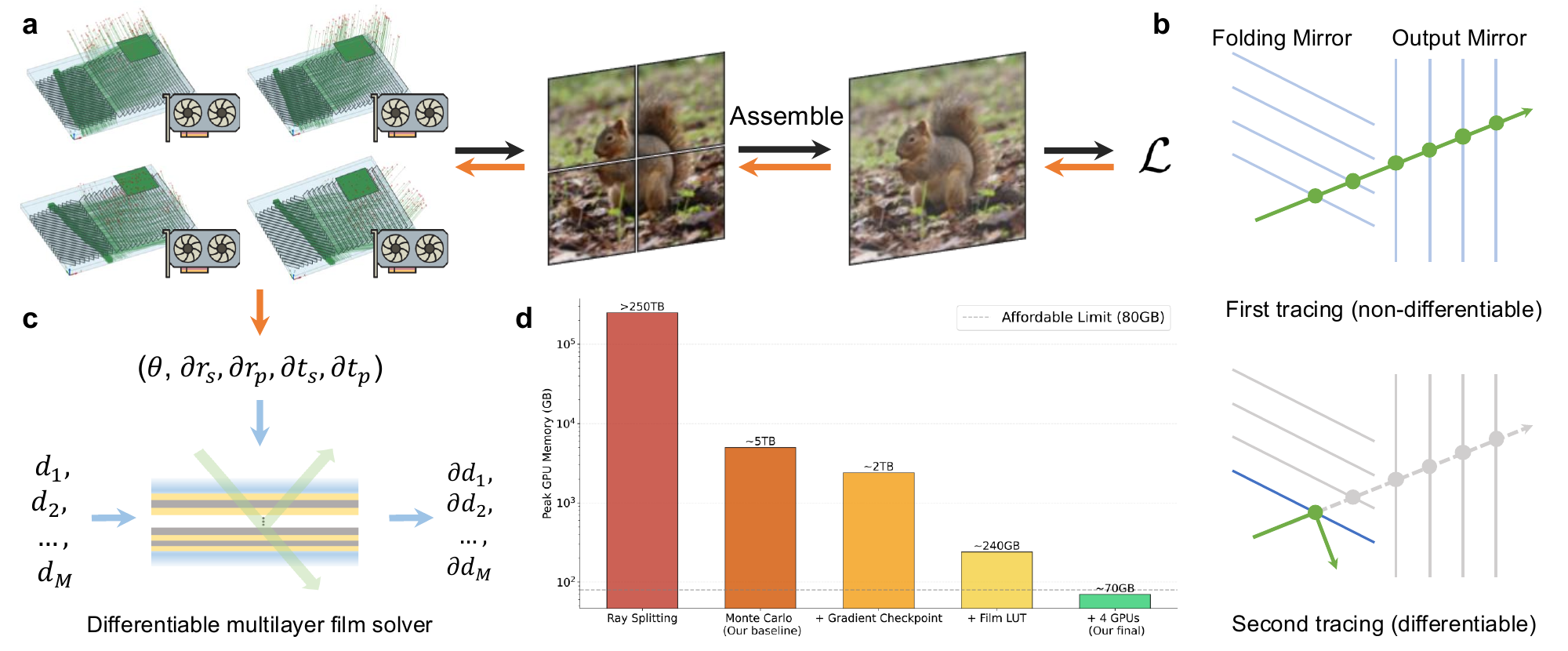}
    \caption{\textbf{Memory-saving strategies for large-scale differentiable optimization of the GWG system.} \textbf{a} The pupil image at the eyebox is partitioned into FoV patches and distributed across multiple GPUs. The patches are assembled into a full-FoV image to compute the loss; in backpropagation, gradients are computed on the full image and then scattered back to the corresponding GPUs. \textbf{b} Differentiable Monte Carlo ray tracing uses a two-pass intersection strategy. In the first pass, we compute ray-surface intersections without tracking gradients and record the intersected surfaces. In the second pass, we recompute only those intersections in differentiable mode. \textbf{c} We use gradient checkpointing to decouple backpropagation through ray tracing and through the thin-film solver. Gradients are first backpropagated to the effective Fresnel coefficients and stored; we then backpropagate through the multilayer solver to update layer thicknesses using stored intermediates. \textbf{d} Peak GPU memory usage with these strategies, enabling large-scale differentiable optimization on a single workstation.}
    \label{fig:methods}
\end{figure}

To make this large-scale optimization tractable, we implement several memory-saving strategies in PyTorch (Fig.~\ref{fig:methods}). We parallelize computation by partitioning the pupil image into FoV patches across GPUs (Fig.~\ref{fig:methods}\textbf{a}), use a two-pass intersection strategy for differentiable non-sequential ray tracing (Fig.~\ref{fig:methods}\textbf{b}), and apply gradient checkpointing to decouple backpropagation through ray tracing and the thin-film solver (Fig.~\ref{fig:methods}\textbf{c}). We additionally precompute the thin-film response on a discrete set of incident angles and use differentiable interpolation to handle intermediate angles. Together, these strategies reduce peak memory usage (Fig.~\ref{fig:methods}\textbf{d}); further details are provided in the Supplementary Note.

\subsection*{Network-Optics Co-design}

Coating optimization primarily corrects global throughput and large-scale non-uniformity, whereas residual artefacts (for example stripe patterns) arise from the discrete PRMA geometry and are difficult to eliminate with coatings alone. We therefore extend the framework to jointly optimize a neural image preprocessor with the GWG coatings (optical-digital co-design) to compensate these residual artefacts. We use the loss
\begin{equation}
    \mathcal{L}' = \|T - \mathcal{N}(T) \odot I\| + \omega \mathcal{L},
    \label{eq:end2end_loss}
\end{equation}
where $T$ is the target displayed image, $I$ is the simulated GWG eyebox response, $\mathcal{N}$ is the neural network, and $\odot$ denotes element-wise multiplication. The weighting factor $\omega$ balances image fidelity and the optical loss. We use NAFNet~\cite{Chen_2022} as a compact backbone (2.68M parameters) to target low-latency deployment. We train from scratch with 1,000 training images and evaluate on 100 validation images. The dataset is captured with a DSLR camera to provide high-resolution natural images (cropped to $1024 \times 1024$) and to avoid copyright constraints. The pipeline is dataset-agnostic and the images are used only to optimize a generic preprocessor.

\begin{figure}[ht]
    \centering
    \includegraphics[width=\textwidth]{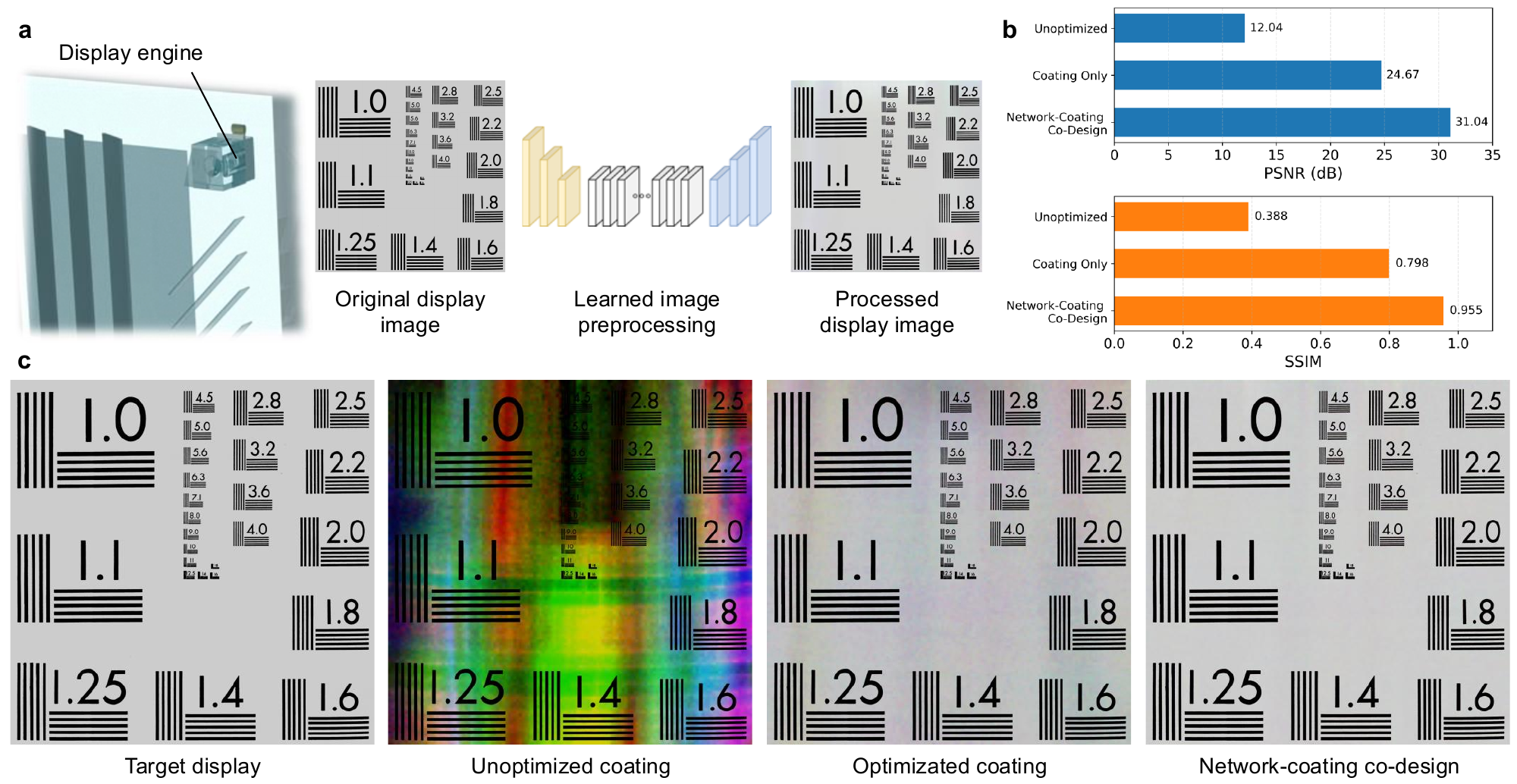}
    \caption{\textbf{End-to-end co-design of image preprocessing and GWG coatings.} \textbf{a} A neural network processes the displayed image before emitted into the GWG. The network and coating parameters are optimized jointly to improve perceived image quality at the eyebox. \textbf{b} PSNR and SSIM measured on validation set. Both coating-only optimization and end-to-end co-design improve image quality, while end-to-end co-design further compensates residual artefacts and improves image quality. \textbf{c} Example outputs. Unoptimized coatings produce dark images with non-uniformities, coating optimization improves brightness and reduces global non-uniformities, and end-to-end co-design further compensates residual artefacts and improves image quality.}
    \label{fig:end2end}
\end{figure}

The images emitted from the display panel are first processed by the neural network before being emitted into the GWG (Fig.~\ref{fig:end2end}\textbf{a}). The network and coating parameters are optimized jointly to improve perceived image quality at the eyebox. On the validation set, PSNR/SSIM improve from 12.04 dB/0.388 (unoptimized) to 24.67 dB/0.798 with coating-only optimization, and to 31.04 dB/0.955 with end-to-end co-design (Fig.~\ref{fig:end2end}\textbf{b}), showing significant improvements in image fidelity with the joint optimization. Example outputs are shown in Fig.~\ref{fig:end2end}\textbf{c}. Coating optimization reduces global non-uniformities, but residual fine-scale artefacts persist due to the waveguide geometry, while the joint optimization further compensates for these artefacts and improves image quality.

\section*{Discussion}

We introduced a large-scale end-to-end differentiable optimization framework for geometric waveguide displays that couples non-sequential light transport with polarization-dependent multilayer thin-film modelling. The approach enables gradient-based optimization of high-dimensional coating stacks in a complex optical system and accelerates convergence relative to sampling-based baselines. By combining probabilistic path sampling, a differentiable transfer-matrix thin-film solver and memory-saving strategies, we can optimize over-parameterized stacks and prune unnecessary layers by driving their thickness towards zero under discrete constraints. In a representative design, these capabilities translate into substantial gains in efficiency and uniformity across the eyebox, and the same framework supports system-level co-design with a neural image preprocessor to further improve perceived image quality.

This end-to-end differentiable formulation shifts GWG coating design from decoupled, mirror-by-mirror tuning to a system-level, gradient-based optimization problem that can be iterated rapidly and explored at high dimensionality. Starting from over-parameterized stacks and pruning layers under discrete constraints reduces the need to pre-specify stack topology, and the same differentiable pipeline enables co-optimization with the display engine (for example, learned image pre-compensation) to target perceptual objectives. However, some pieces are missing for deployment, for example, experimental validation.

Several limitations motivate future improvements. First, we primarily optimized coating thicknesses with a fixed PRMA geometry. Extending the parameterization to include mirror tilts and rotations would provide additional degrees of freedom to shape energy flow and could further improve efficiency and uniformity. Second, our objective targets eyebox image quality under the assumed system model. Incorporating additional physical factors relevant to real-world deployment, for example stray light, waveguide leakage and environmental reflections, would allow the optimizer to explicitly trade off see-through quality and artefact suppression. Finally, although we validate optimized designs with an independent forward simulation workflow (Supplementary Note), experimental prototypes will be an important step towards deployment.

More broadly, the differentiable non-sequential framework opens several research directions. Supporting curved waveguide geometries by parameterizing substrate curvature would enable co-optimization for ergonomics and aesthetics in consumer form factors~\cite{Draper_2022,Weng_2025}. Integrating more realistic projection optics (including aberrations, alignment tolerances and coupling efficiencies) would move towards true end-to-end optimization from the light source to the eyebox. Beyond waveguides, the same combination of differentiable ray tracing and thin-film modelling could be applied to anti-ghosting coating design for refractive optics or lens coating design by incorporating ghost-path analysis directly into the loss.
\section*{Methods}

\subsection*{Multilayer thin-film solver}

The polarization of reflected and transmitted rays is modulated by multilayer coatings on each mirror. We use the Fresnel coefficients at an interface between medium $i$ and medium $j$. For s-polarization (TE) and p-polarization (TM), these coefficients are:

\begin{equation}
    \begin{aligned}
        r_s &= \frac{n_i \cos \theta_i - n_j \cos \theta_t}{n_i \cos \theta_i + n_j \cos \theta_t}, \quad
        &t_s &= \frac{2 n_i \cos \theta_i}{n_i \cos \theta_i + n_j \cos \theta_t} \\
        r_p &= \frac{n_j \cos \theta_i - n_i \cos \theta_t}{n_j \cos \theta_i + n_i \cos \theta_t}, \quad
        &t_p &= \frac{2 n_i \cos \theta_i}{n_j \cos \theta_i + n_i \cos \theta_t}
    \end{aligned}
    \label{eq:fresnel_coefficients}
\end{equation}

\noindent where $n_i$ and $n_j$ are the refractive indices of media $i$ and $j$, $\theta_i$ is the incident angle, and $\theta_t$ is the transmitted angle determined by Snell's law. To model thin films, we compute the effective complex reflection ($r_{eq}$) and transmission ($t_{eq}$) coefficients for a single layer by summing the amplitudes of multiple internal reflections (Airy formulas):

\begin{equation}
    r_{eq} = \frac{r_{12} + r_{23}e^{-2i\delta}}{1 + r_{12}r_{23}e^{-2i\delta}}, \quad
    t_{eq} = \frac{t_{12}t_{23}e^{-i\delta}}{1 + r_{12}r_{23}e^{-2i\delta}}
    \label{eq:airy_formulas}
\end{equation}

\noindent where $r_{12}, t_{12}$ and $r_{23}, t_{23}$ are the Fresnel coefficients at the two layer boundaries. The phase thickness is $\delta = 2\pi n d \cos \theta / \lambda$, where $d$ is the layer thickness, $n$ is the refractive index, $\lambda$ is the wavelength and $\theta$ is the propagation angle inside the layer. To obtain the effective reflection and transmission coefficients of a multilayer stack (for both s and p polarizations), we use the transfer matrix method (TMM)~\cite{heavens1960optical}. Accuracy and efficiency evaluations are provided in the Supplementary Note.

\subsection*{Polarization ray tracing}

We perform polarization ray tracing~\cite{chipman1995mechanics} to propagate the complex electric-field vector of each ray. At each coated interface, we project the field onto the local s and p bases. Using the effective Fresnel coefficients computed above, the reflected and transmitted complex amplitudes are
\begin{equation}
    \begin{aligned}
        \mathbf{E}_{\text{reflected}} = r_s \mathbf{E}_s \hat{s} + r_p \mathbf{E}_p \hat{p}, \quad
        \mathbf{E}_{\text{transmitted}} = t_s \mathbf{E}_s \hat{s}' + t_p \mathbf{E}_p \hat{p}',
    \end{aligned}  
    \label{eq:complex_amplitudes}
\end{equation}

\noindent where $\hat{s}$ and $\hat{p}$ are the local s and p unit vectors for the reflected ray, and $\hat{s}'$ and $\hat{p}'$ are the corresponding unit vectors for the transmitted ray. The coefficients $(r_s, r_p, t_s, t_p)$ are the effective Fresnel coefficients computed in the previous section. Because the s/p bases depend on the surface normal and ray direction, we recompute the basis and re-project the field at every interface before applying the transform.

Equations~\eqref{eq:fresnel_coefficients}, \eqref{eq:airy_formulas} and~\eqref{eq:complex_amplitudes} relate coating performance to the design parameters (layer thicknesses), incident angle and polarization state. Implemented in PyTorch, the solver supports automatic differentiation, enabling gradient-based optimization of layer thickness. In our experiments, we use SiO$_2$ ($n=1.46$) for the first and last layers and Ta$_2$O$_5$ ($n=2.13$) for intermediate layers; the glass substrate has refractive index $n=1.9$. Each layer thickness is constrained to be between 20~nm and 200~nm. Further implementation details are provided in the Supplementary Note.

\subsection*{Differentiable Monte Carlo non-sequential ray tracing}

Conventional non-sequential ray tracing splits a ray into reflected and transmitted branches at each interaction~\cite{zemax_2024}. This leads to exponential growth in ray count and high computational and memory costs for complex GWG architectures. It is also difficult to differentiate efficiently, as a fully differentiable implementation would further increase memory and compute. We instead use differentiable Monte Carlo non-sequential ray tracing. At each partially reflective mirror, a ray is stochastically reflected or transmitted. To preserve energy in polarization ray tracing, we scale the complex amplitude of the sampled path by the sampling probability:
\begin{equation}
    \mathbf{E}'_{\text{reflected}} = \frac{\mathbf{E}_{\text{reflected}}}{\sqrt{\omega}}, \quad
    \mathbf{E}'_{\text{transmitted}} = \frac{\mathbf{E}_{\text{transmitted}}}{\sqrt{1 - \omega}}
\end{equation}
\noindent where $\omega$ is the reflection probability at the mirror. Monte Carlo sampling keeps the number of rays constant, enabling efficient GPU parallelism with bounded memory. To enable gradients through stochastic sampling, we use a reparameterization that decouples event sampling from the physical coefficients, allowing gradients to propagate in the backward pass. To further reduce memory, we adopt the two-pass intersection strategy described above (Fig.~\ref{fig:methods}b), which limits the autodiff graph to the surfaces actually intersected by each ray.

Similar to differentiable sequential ray tracing~\cite{Wang_2022,Yang_2024,yang2024hybrid}, this formulation enables gradient-based optimization while maintaining efficient GPU parallelism. In our implementation, the bounded memory footprint enables large-scale simulations for GWG optimization on a single workstation equipped with four NVIDIA A100 (80GB) GPUs. The intensity at the eyebox is computed as the incoherent sum of squared complex amplitudes, ${\rm I} = \sum_i |\mathbf{E}_i|^2$. The reparameterization enables gradient backpropagation from the loss (Eq.~\ref{eq:loss}) through ray tracing to the coating thickness parameters. Additional implementation details are provided in the Supplementary Note.
 
\subsection*{Learned reflection probabilities}

To improve sampling efficiency and optimization stability, we use learned reflection probabilities ($\omega$) during optimization. We first run a pre-optimization stage to determine $\omega$ for each mirror, aiming to maximize throughput to the pupil plane. In this stage, each coating is idealized as a scalar reflectance (fraction of incident energy reflected) and we perform geometric ray tracing without polarization. We then optimize the multilayer thicknesses using polarization tracing while using the learned $\omega$ for Monte Carlo sampling. Without this pre-optimization, few rays reach the pupil plane, leading to unstable optimization and noisy gradients. Further details are provided in the Supplementary Note.

\bibliographystyle{unsrt}  
\bibliography{references}

@article{heavens1960optical,
  title     = {Optical properties of thin films},
  author    = {Heavens, Oskar Sigmund},
  journal   = {Reports on Progress in Physics},
  volume    = {23},
  number    = {1},
  pages     = {1},
  year      = {1960},
  publisher = {IOP Publishing}
}

@article{chipman1995mechanics,
  title     = {Mechanics of polarization ray tracing},
  author    = {Chipman, Russell A},
  journal   = {Optical Engineering},
  volume    = {34},
  number    = {6},
  pages     = {1636--1645},
  year      = {1995},
  publisher = {SPIE}
}

@article{Moharam_1982,
  title     = {Diffraction analysis of dielectric surface-relief gratings},
  volume    = {72},
  issn      = {0030-3941},
  url       = {http://dx.doi.org/10.1364/josa.72.001385},
  doi       = {10.1364/josa.72.001385},
  number    = {10},
  journal   = {Journal of the Optical Society of America},
  publisher = {Optica Publishing Group},
  author    = {Moharam, M. G. and Gaylord, T. K.},
  year      = {1982},
  month     = oct,
  pages     = {1385}
}

@article{Miller_1997,
  title     = {Design and fabrication of binary slanted surface-relief gratings for a planar optical interconnection},
  volume    = {36},
  issn      = {1539-4522},
  url       = {http://dx.doi.org/10.1364/ao.36.005717},
  doi       = {10.1364/ao.36.005717},
  number    = {23},
  journal   = {Applied Optics},
  publisher = {Optica Publishing Group},
  author    = {Miller, J. Michael and de Beaucoudrey, Nicole and Chavel, Pierre and Turunen, Jari and Cambril, Edmond},
  year      = {1997},
  month     = aug,
  pages     = {5717}
}

@article{rolland2000wide,
  title     = {Wide-angle, off-axis, see-through head-mounted display},
  author    = {Rolland, Jannick P},
  journal   = {Optical engineering},
  volume    = {39},
  number    = {7},
  pages     = {1760--1767},
  year      = {2000},
  publisher = {SPIE}
}

@article{kress2013review,
  title     = {A review of head-mounted displays (HMD) technologies and applications for consumer electronics},
  author    = {Kress, Bernard and Starner, Thad},
  journal   = {Photonic Applications for Aerospace, Commercial, and Harsh Environments IV},
  volume    = {8720},
  pages     = {62--74},
  year      = {2013},
  publisher = {SPIE}
}

@article{hua20143d,
  title     = {A 3D integral imaging optical see-through head-mounted display},
  author    = {Hua, Hong and Javidi, Bahram},
  journal   = {Optics express},
  volume    = {22},
  number    = {11},
  pages     = {13484--13491},
  year      = {2014},
  publisher = {Optical Society of America}
}

@article{Jang_2017,
  title     = {Retinal 3D: augmented reality near-eye display via pupil-tracked light field projection on retina},
  volume    = {36},
  issn      = {1557-7368},
  url       = {http://dx.doi.org/10.1145/3130800.3130889},
  doi       = {10.1145/3130800.3130889},
  number    = {6},
  journal   = {ACM Transactions on Graphics},
  publisher = {Association for Computing Machinery (ACM)},
  author    = {Jang, Changwon and Bang, Kiseung and Moon, Seokil and Kim, Jonghyun and Lee, Seungjae and Lee, Byoungho},
  year      = {2017},
  month     = nov,
  pages     = {1–13}
}

@article{Lee_2018,
  title     = {Metasurface eyepiece for augmented reality},
  volume    = {9},
  issn      = {2041-1723},
  url       = {http://dx.doi.org/10.1038/s41467-018-07011-5},
  doi       = {10.1038/s41467-018-07011-5},
  number    = {1},
  journal   = {Nature Communications},
  publisher = {Springer Science and Business Media LLC},
  author    = {Lee, Gun-Yeal and Hong, Jong-Young and Hwang, SoonHyoung and Moon, Seokil and Kang, Hyeokjung and Jeon, Sohee and Kim, Hwi and Jeong, Jun-Ho and Lee, Byoungho},
  year      = {2018},
  month     = nov
}

@article{gu2018design,
  title     = {Design of a two-dimensional stray-light-free geometrical waveguide head-up display},
  author    = {Gu, Luo and Cheng, Dewen and Wang, Qiwei and Hou, Qichao and Wang, Yongtian},
  journal   = {Applied Optics},
  volume    = {57},
  number    = {31},
  pages     = {9246--9256},
  year      = {2018},
  publisher = {Optical Society of America}
}

@article{Xu_2019,
  title     = {Methods of optimizing and evaluating geometrical lightguides with microstructure mirrors for augmented reality displays},
  volume    = {27},
  issn      = {1094-4087},
  url       = {http://dx.doi.org/10.1364/oe.27.005523},
  doi       = {10.1364/oe.27.005523},
  number    = {4},
  journal   = {Optics Express},
  publisher = {The Optical Society},
  author    = {Xu, Miaomiao and Hua, Hong},
  year      = {2019},
  month     = feb,
  pages     = {5523}
}

@article{Wang_2022,
  title     = {dO: A Differentiable Engine for Deep Lens Design of Computational Imaging Systems},
  volume    = {8},
  issn      = {2573-0436},
  url       = {http://dx.doi.org/10.1109/tci.2022.3212837},
  doi       = {10.1109/tci.2022.3212837},
  journal   = {IEEE Transactions on Computational Imaging},
  publisher = {Institute of Electrical and Electronics Engineers (IEEE)},
  author    = {Wang, Congli and Chen, Ni and Heidrich, Wolfgang},
  year      = {2022},
  pages     = {905–916}
}

@article{lee2020foveated,
  title     = {Foveated near-eye display for mixed reality using liquid crystal photonics},
  author    = {Lee, Seungjae and Wang, Mengfei and Li, Gang and Lu, Lu and Sulai, Yusufu and Jang, Changwon and Silverstein, Barry},
  journal   = {Scientific Reports},
  volume    = {10},
  number    = {1},
  pages     = {16127},
  year      = {2020},
  publisher = {Nature Publishing Group UK London}
}

@patent{Danziger2021,
  author   = {Yochay Danziger and Ronen Chiki and Jonathan Gelberg},
  title    = {Optical systems including light-guide optical elements with two-dimensional expansion},
  number   = {US10983264B2},
  year     = {2021},
  assignee = {Lumus Ltd},
  month    = {April},
  url      = {https://patents.google.com/patent/US10983264B2/en}
}

@article{Xiong_2021,
  title     = {Augmented reality and virtual reality displays: emerging technologies and future perspectives},
  volume    = {10},
  issn      = {2047-7538},
  url       = {http://dx.doi.org/10.1038/s41377-021-00658-8},
  doi       = {10.1038/s41377-021-00658-8},
  number    = {1},
  journal   = {Light: Science \& Applications},
  publisher = {Springer Science and Business Media LLC},
  author    = {Xiong, Jianghao and Hsiang, En-Lin and He, Ziqian and Zhan, Tao and Wu, Shin-Tson},
  year      = {2021},
  month     = oct
}

@article{Cheng_2022,
  title     = {Design method of a wide-angle AR display with a single-layer two-dimensional pupil expansion geometrical waveguide},
  volume    = {61},
  issn      = {2155-3165},
  url       = {http://dx.doi.org/10.1364/ao.459644},
  doi       = {10.1364/ao.459644},
  number    = {19},
  journal   = {Applied Optics},
  publisher = {Optica Publishing Group},
  author    = {Cheng, Dewen and Wang, Qiwei and Wei, Li and Wang, Ximeng and Zhou, Lijun and Hou, Qichao and Duan, Jiaxi and Yang, Tong and Wang, Yongtian},
  year      = {2022},
  month     = jun,
  pages     = {5813}
}

@article{Ni_2022,
  title     = {Uniformity improvement of two-dimensional surface relief grating waveguide display using particle swarm optimization},
  volume    = {30},
  issn      = {1094-4087},
  url       = {http://dx.doi.org/10.1364/oe.462384},
  doi       = {10.1364/oe.462384},
  number    = {14},
  journal   = {Optics Express},
  publisher = {Optica Publishing Group},
  author    = {Ni, Dongwei and Cheng, Dewen and Liu, Yue and Wang, Ximeng and Yao, Cheng and Yang, Tong and Chi, Cheng and Wang, Yongtian},
  year      = {2022},
  month     = jun,
  pages     = {24523}
}

@article{Draper_2022,
  title     = {Holographic curved waveguide combiner for HUD/AR with 1-D pupil expansion},
  volume    = {30},
  issn      = {1094-4087},
  url       = {http://dx.doi.org/10.1364/oe.445091},
  doi       = {10.1364/oe.445091},
  number    = {2},
  journal   = {Optics Express},
  publisher = {Optica Publishing Group},
  author    = {Draper, Craig T. and Blanche, Pierre-Alexandre},
  year      = {2022},
  month     = jan,
  pages     = {2503}
}

@inbook{Chen_2022,
  title     = {Simple Baselines for Image Restoration},
  isbn      = {9783031200717},
  issn      = {1611-3349},
  url       = {http://dx.doi.org/10.1007/978-3-031-20071-7_2},
  doi       = {10.1007/978-3-031-20071-7_2},
  booktitle = {Computer Vision – ECCV 2022},
  publisher = {Springer Nature Switzerland},
  author    = {Chen, Liangyu and Chu, Xiaojie and Zhang, Xiangyu and Sun, Jian},
  year      = {2022},
  pages     = {17–33}
}

@article{Ruan_2023,
  title     = {Design method of an ultra-thin two-dimensional geometrical waveguide near-eye display based on forward-ray-tracing and maximum FOV analysis},
  volume    = {31},
  issn      = {1094-4087},
  url       = {http://dx.doi.org/10.1364/oe.498011},
  doi       = {10.1364/oe.498011},
  number    = {21},
  journal   = {Optics Express},
  publisher = {Optica Publishing Group},
  author    = {Ruan, Ningye and Shi, Feng and Tian, Ye and Xing, Peng and Zhang, Wanli and Qiao, Shuo},
  year      = {2023},
  month     = sep,
  pages     = {33799}
}

@article{weng2023high,
  title     = {High-efficiency and compact two-dimensional exit pupil expansion design for diffractive waveguide based on polarization volume grating},
  author    = {Weng, Yishi and Zhang, Yuning and Wang, Wei and Gu, Yuchen and Wang, Chuang and Wei, Ran and Zhang, Lixuan and Wang, Baoping},
  journal   = {Optics Express},
  volume    = {31},
  number    = {4},
  pages     = {6601--6614},
  year      = {2023},
  publisher = {Optica Publishing Group}
}

@article{Tseng_2024,
  title     = {Neural étendue expander for ultra-wide-angle high-fidelity holographic display},
  volume    = {15},
  issn      = {2041-1723},
  url       = {http://dx.doi.org/10.1038/s41467-024-46915-3},
  doi       = {10.1038/s41467-024-46915-3},
  number    = {1},
  journal   = {Nature Communications},
  publisher = {Springer Science and Business Media LLC},
  author    = {Tseng, Ethan and Kuo, Grace and Baek, Seung-Hwan and Matsuda, Nathan and Maimone, Andrew and Schiffers, Florian and Chakravarthula, Praneeth and Fu, Qiang and Heidrich, Wolfgang and Lanman, Douglas and Heide, Felix},
  year      = {2024},
  month     = apr
}

@article{rolland2024waveguide,
  title     = {Waveguide-based augmented reality displays: a highlight},
  author    = {Rolland, Jannick P and Goodsell, Jeremy},
  journal   = {Light: Science \& Applications},
  volume    = {13},
  number    = {1},
  pages     = {22},
  year      = {2024},
  publisher = {Nature Publishing Group UK London}
}

@inproceedings{yang2024hybrid,
  title     = {End-to-end hybrid refractive-diffractive lens design with differentiable ray-wave model},
  author    = {Yang, Xinge and Souza, Matheus and Wang, Kunyi and Chakravarthula, Praneeth and Fu, Qiang and Heidrich, Wolfgang},
  booktitle = {SIGGRAPH Asia 2024 Conference Papers},
  pages     = {1--11},
  year      = {2024}
}

@article{Gopakumar_2024,
  title     = {Full-colour 3D holographic augmented-reality displays with metasurface waveguides},
  volume    = {629},
  issn      = {1476-4687},
  url       = {http://dx.doi.org/10.1038/s41586-024-07386-0},
  doi       = {10.1038/s41586-024-07386-0},
  number    = {8013},
  journal   = {Nature},
  publisher = {Springer Science and Business Media LLC},
  author    = {Gopakumar, Manu and Lee, Gun-Yeal and Choi, Suyeon and Chao, Brian and Peng, Yifan and Kim, Jonghyun and Wetzstein, Gordon},
  year      = {2024},
  month     = may,
  pages     = {791–797}
}

@article{Yang_2024,
  title     = {Curriculum learning for ab initio deep learned refractive optics},
  volume    = {15},
  issn      = {2041-1723},
  url       = {http://dx.doi.org/10.1038/s41467-024-50835-7},
  doi       = {10.1038/s41467-024-50835-7},
  number    = {1},
  journal   = {Nature Communications},
  publisher = {Springer Science and Business Media LLC},
  author    = {Yang, Xinge and Fu, Qiang and Heidrich, Wolfgang},
  year      = {2024},
  month     = aug
}

@article{Jang_2024,
  title     = {Waveguide holography for 3D augmented reality glasses},
  volume    = {15},
  issn      = {2041-1723},
  url       = {http://dx.doi.org/10.1038/s41467-023-44032-1},
  doi       = {10.1038/s41467-023-44032-1},
  number    = {1},
  journal   = {Nature Communications},
  publisher = {Springer Science and Business Media LLC},
  author    = {Jang, Changwon and Bang, Kiseung and Chae, Minseok and Lee, Byoungho and Lanman, Douglas},
  year      = {2024},
  month     = jan
}

@book{zemax_2024,
  title     = {Designing Optics Using Zemax OpticStudio®},
  isbn      = {9781510668577},
  url       = {http://dx.doi.org/10.1117/3.100004},
  doi       = {10.1117/3.100004},
  publisher = {SPIE},
  author    = {O’Shea, Donald C. and Bentley, Julie L.},
  year      = {2024},
  month     = jan
}

@article{Weng_2025,
  title     = {Design and fabrication of curved waveguide display based on freeform polarization volume holograms},
  volume    = {33},
  issn      = {1094-4087},
  url       = {http://dx.doi.org/10.1364/oe.557298},
  doi       = {10.1364/oe.557298},
  number    = {7},
  journal   = {Optics Express},
  publisher = {Optica Publishing Group},
  author    = {Weng, Jiacheng and Pei, Chunyang and Li, Haifeng and Wu, Rengmao and Liu, Xu},
  year      = {2025},
  month     = mar,
  pages     = {15362}
}

@article{shi2025flat,
  title     = {Flat-panel laser displays through large-scale photonic integrated circuits},
  author    = {Shi, Zhujun and Cheng, Risheng and Wei, Guohua and Hickman, Steven A and Shin, Min Chul and Topalian, Peter and Wang, Lei and Coso, Dusan and Wang, Youmin and Wang, Qingjun and others},
  journal   = {Nature},
  volume    = {644},
  number    = {8077},
  pages     = {652--659},
  year      = {2025},
  publisher = {Nature Publishing Group UK London}
}

@article{choi2025synthetic,
  title={Synthetic aperture waveguide holography for compact mixed-reality displays with large {\'e}tendue},
  author={Choi, Suyeon and Jang, Changwon and Lanman, Douglas and Wetzstein, Gordon},
  journal={Nature Photonics},
  volume={19},
  number={8},
  pages={854--863},
  year={2025},
  publisher={Nature Publishing Group UK London}
}

\end{document}